\documentclass[aps,floatfix,twocolumn,showpacs]{revtex4}
\usepackage{natbib}
\usepackage{graphicx}
\usepackage{graphics,epsfig}
\usepackage{amssymb}

\newcommand{\physrep}{Phys.~Rep.}
\newcommand{\araa}{Ann.~Rev.~Astron.~Astrophys.}

\def\be {\begin{equation}}
\def\ee {\end{equation}}
\def\bea {\begin{eqnarray}}
\def\eea {\end{eqnarray}}

\def \lf {\left(}
\def \rg {\right)}
\begin{document}
\title{Dark Energy or Apparent Acceleration Due to a Relativistic Cosmological Model More Complex than FLRW?}
\author{Mustapha Ishak$^1$\footnote{Electronic address: mishak@utdallas.edu}, James Richardson$^1$, David Garred$^1$, Delilah Whittington$^1$, Anthony Nwankwo$^1$, Roberto Sussman$^2$} 
\affiliation{
$^1$ Department of Physics, University of Texas at Dallas, Richardson, TX 75083, USA\\
$^2$ ICN Institute, Universidad Nacional Aut\'onoma de M\'exico, 04510 D.F., M\'exico.
}
\date{\today}
\date{\today}
\begin{abstract}
We use the Szekeres inhomogeneous relativistic models in order to fit supernova combined data sets. We show that with a choice of the spatial curvature function that is guided by current observations, the models fit the supernova data almost as well as the LCDM model without requiring a dark energy component. The Szekeres models were originally derived as an exact solution to Einstein's equations with a general metric that has no symmetries and are regarded as good candidates to model the true lumpy universe that we observe. The null geodesics in these models are not radial. The best fit model found is also consistent with the requirement of spatial flatness at CMB scales. The first results presented here seem to encourage further investigations of apparent acceleration using various inhomogeneous models and other constraints from CMB and large structure need to be explored next.
\end{abstract}
\pacs{98.80.Es,95.36.+x,98.80.-k}
\maketitle
%
 
\section{Introduction.}
\label{sec:introduction}
Complementary cosmological observations have established that the expansion of the universe has entered a phase of acceleration \cite{observations1,observations2,observations3,observations4,observations5,observations7,observations8,observations9}. Cosmic acceleration and the dark energy associated with it have been recognized as one of the most important and challenging current problems in cosmology and physics, see e.g. the reviews \cite{reviews1,reviews2,reviews3,reviews4,reviews5,reviews6,reviews7}. As discussed in the literature, cosmic acceleration can be caused by a repulsive dark energy, or, can be caused by some radical changes to gravity theory (General Relativity). However, a third possibility has been discussed in a number of recent papers and that is cosmic acceleration could be an apparent effect because cosmological observables, such as the luminosity-distance to supernovae, are altered by inhomogeneities in the universe when analyzed within exact inhomogeneous models. In view of the challenges presented by the cosmic acceleration problem, all the three possibilities need to be thoroughly explored.

The possibility of an apparent acceleration has been discussed in the literature using the inhomogeneous Lemaitre-Tolman-Bondi (LTB) models, an incomplete list includes \cite{Iguchietal2002,TanimotoNambu2007, Mansouri2005,Alnesetal2006,Garfinkle2006,Kaietal2006,Biswasetal2006}. In these and other papers, it was shown that the LTB models can fit supernova observations and the position of the first peak of the Cosmic Microwave Background Radiation (CMB) without any dark energy in the models. Although the LTB models are spherically symmetric, the results of those papers support, as a proof of concept, that apparent acceleration is a serious possibility. Thus, it is a compelling endeavor to explore apparent acceleration within more general inhomogeneous cosmological models. 

In this paper, we present a first analysis of apparent acceleration using the appealing Szekeres inhomogeneous models \cite{Szekeres1,Szekeres2}; see also \cite{krasinski1997} and references therein for a review of the models. The models were originally derived by Szekeres \cite{Szekeres1,Szekeres2} as an exact solution to Einstein's equations with a general metric that has no symmetries (i.e. no Killing vector fields). The models have been investigated analytically by several authors \cite{Szafron1977, BonnorAndTomimura1976,Bonnoretal1977,GoodeAndWainwright1982,Bonnor1985,SussmanAndTriginer1999,HellabyAndKrasinski2002,Bolejko2006a,Bolejko2006b,Bolejko2007} and are regarded as the best exact solution candidates to represent the true lumpy universe we live in. For example, the models are put in the same classification as the observed lumpy universe in \cite{EllisVanEllst1998} (see the table on page 37 there). In reference \cite{GoodeAndWainwright1982}, the authors reformulated the models using a coordinate system that shows that the models can be regarded as non-linear exact perturbations of the Friedmann-Lemaitre-Robertson-Walker (FLRW) models but where the exact non-linearity of General Relativity is not altered. As we discuss further, the Szekeres models have a flexible geometrical structure and are very likely to fit various cosmological data sets without the need for a dark energy component. 

It is perhaps worth clarifying that we are not proposing the Szekeres model as the true model of the universe but rather investigating the possibility of an apparent acceleration in an inhomogeneous universe. 

In this scenario, apparent acceleration is due to the fact that we happen to live in one of the many under-dense regions of the universe. The physical reason for such an observation is that inside the under-dense region, there is less matter to slow down the expansion and therefore the expansion rate in the under-dense region is larger than what it is in an over-dense region or compared to the overall averaged rate of expansion. Some limits have been put on the dispersion in this expansion, see for example \cite{Li}, but this was contested, see for example \cite{Sarkar,Hunt,Alexander}. So the difference between the dynamics inside the under-dense region compared to outside of it could give the apparent effect of an accelerating expansion. Furthermore, in these models we do not have to impose that we are located close to the center of the under-dense region because the definition of a center itself depends on a spherically symmetric geometry which our models do not suffer from. 

Interestingly, the possibility of apparent acceleration connects to the averaging problem in relativity and cosmology, e.g. \cite{Ellis1984,Ellisetal1985,Matravers1995}. The problem states that the operation of smoothing inhomogeneities in the universe from small scales of distance to large scales is an operation that does not commute with the operation of applying the Einstein equations. In other words, the order in which the two operations are applied is important and leads to different results due to the non-linearity of Einstein's equations. Because of this non-commutation problem, it becomes clear why using observables in inhomogeneous models, such as the Szekeres models, is different from using FLRW plus perturbations and why it can lead to different interpretations. The problem was pointed out well before the cosmic acceleration problem and has now re-emerged because of dark energy and possible links to inhomogeneous models; for a recent review see \cite{Celerier2007} and references therein. In our work here, non-linearities affect observables and lead to apparent acceleration, and this is different from other ideas where authors tried to use non-linearities in order to generate a true acceleration; see also \cite{Celerier2007} for a review. 

Finally, it is worth mentioning that the possibility of apparent acceleration also addresses the acceleration/dark energy coincidence problem, e.g. reviews \cite{reviews1,reviews2,reviews3,reviews4,reviews5,reviews6,reviews7}. This problem is stated in literature as the "why now?" problem, i.e. why cosmic acceleration is taking place during the current epoch of cosmic evolution and only after initial structures had formed? As pointed out in previous literature, e.g. \cite{Celerier2007} and references therein, in the apparent acceleration scenario, the effect is taking place at the current epoch because of the onset of non-linear cosmic structures in the universe. 

The usual difficulty in comparing inhomogeneous models with observations is that it is not possible, in general, to analytically solve the null geodesic equation in these models and derive observable functions. This equation is easily solved in the FLRW model, but that is not the case in more complex models. In the Szekeres models, unlike the LTB and the FLRW models, the null geodesics are not radial so the full set of equations need to be integrated in order to consider observations. In the present work, we explore a numerical integration in order to overcome the problem. 

In this first of a series of papers, we launch a program where we plan to use relativistic cosmological models that are more complex than the FLRW models (focusing on the Szekeres and Szekeres-Szafron models) and compare them to currently available and future cosmological data. We show in this paper that the Szekeres models successfully fit supernova data and provide preliminary support to the possibility of apparent acceleration. A number of points need to be further investigated and are subject to ongoing and future work by the authors. 
%
 
\section{The Szekeres Cosmological Models} 
\label{sec:szekeres}

The models are described in some detail in \cite{krasinski1997,PlebanskiAndKrasinski2006} and we give here a brief overview only. We use the set of coordinates used by \cite{Bolejko2006a,Bolejko2007} and by \cite{HellabyAndKrasinski2002}. The spacetime metric reads  
\be
ds^2= -dt^2+\frac{(R'-R\frac{E'}{E})^2}{\epsilon-k(r)}dr^2 +R^2\frac{dp^2+dq^2}{E^2}
\label{eq:metric}
\ee
where $R=R(t,r)$ is the area distance and $'$ is used for $\partial/\partial r$. The function $k(r)$ is related to the energy per unit mass and determines the curvature of the spatial sections $t=$constant. This function divides the models into sub-cases: hyperbolic $(k(r)<0)$, parabolic $(k(r)=0)$, and elliptic $(k(r)>0)$. The function $E=E(r,p,q)$ is given by
\be
E(r,p,q)=\frac{(p^2+q^2)}{2S(r)}-\frac{P(r)}{S(r)}p-\frac{Q(r)}{S(r)}q+C(r)
\ee
and the functions $P(r),S(r),Q(r)$ and $C(r)$ satisfy the relation 
\be
C(r)=\frac{P^2(r)}{2S(r)}+\frac{Q^2(r)}{2S(r)}+\frac{S(r)}{2}\,\epsilon, 
\ee
but are otherwise arbitrary. The geometrical function $E(r,p,q)$ and the geometrical constant $\epsilon=0,+1$,or $-1$ define further sub-cases of the Szekeres models and control the mapping of their various hyper-surfaces \cite{HellabyAndKrasinski2002}.  

The Einstein field equations with a dust source are 
\be
\dot{R}^2(t,r)=\frac{2M(r)}{R(t,r)}-k(r)
\label{eq:EE1}
\ee
and 
\be
8 \pi \rho(t,r,p,q)=\frac{2(M'(r)-3 M(r) \frac{E'}{E})}{R^2(R'-R \frac{E'}{E})}
\label{eq:EE2}
\ee
where the function $M(r)$ represents the total active gravitational mass in the case $\epsilon =+1$ \cite{HellabyAndKrasinski2002,HellabyAndKrasinski2008}. The evolution of $R(t,r)$ depends on $k(r)$ and is given by:

\noindent
Hyperbolic case: $k(r)<0$
\be
R(t,r)=\frac{M(r)}{-k(r)}(\cosh{\eta}-1)
\ee
\be
t-t_B(r)=\frac{M(r)}{-k(r)^{3/2}}(\sinh{\eta}-\eta)
\ee
Parabolic case: $k(r)=0$
\be
R(t,r)=M(r) \frac{\eta^2}{2}
\ee
\be
t-t_B(r)=M(r)\frac{\eta^3}{6}
\ee
Elliptic case: $k(r)>0$
\be
R(t,r)=\frac{M(r)}{k(r)}(1-\cos{\eta})
\ee
\be
t-t_B(r)=\frac{M(r)}{(k(r))^{3/2}}(\eta-\sin{\eta})
\ee
where $t_B(r)$ is an arbitrary function of $r$ and represents the Big Bang time. For a simultaneous Big Bang, we choose $t_B=constant= 0$ just as in an FLRW model. It can be seen that there are several sub-cases of the Szekeres models and that a given model is specified by six functions that can be reduced to five by using the coordinate freedom in $r$. The model we chose for our analysis is specified in section IV.

\section{The null geodesic equations in the Szekeres models}
The null geodesic equation describes the motion of light rays arriving to us from astronomical objects, and it is necessary to solve it in order to derive observables, such as the luminosity-distance to supernovae or the angular diameter distance to the CMB last scattering surface. This equation is easily solved in the FLRW but not in the Szekeres models, and here we employ a numerical approach to the problem. 

Furthermore, it is worth noting that unlike in the LTB geometry, there are no radial null geodesics in the Szekeres models as, for example, was discussed in \cite{PlebanskiAndKrasinski2006,Nolan2007,Bolejko2008}. This is due to the dependence of these geodesics on the $p$ and $q$ coordinates and one must integrate the full set of geodesic equations as we proceed below. 

First, we apply the null geodesic equation 
\be
k_{a;b}k^b=0
\label{eq:SzekNG}  
\ee
\noindent using the Szekeres metric (\ref{eq:metric}) and where $k^a$ is a null vector ($k_ak^a=0$) tangent to the geodesics. The resulting equations for the coordinates $\{t, r, p, q\}$ are written below. These were previously given in for example \cite{Nolan2007,Bolejko2008} and we rederived them here using the computer algebra system GRTensorII \cite{GRTensorII}: 
\bea
\label{eq:NGt}
&&\frac{d^2t}{d\lambda^2} + \frac{R_{,tr}-R_{,t}E_{,r}}{1-k}\left(R_{,r}-R\frac{E_{,r}}{E}\right)(\frac{dr}{d\lambda})^2 \\ \nonumber 
&&+\frac{RR_{,t}}{E^2}\lf(\frac{dp}{d\lambda})^2+(\frac{dq}{d\lambda})^2\rg = 0 
\eea
\bea
\label{eq:NGr} 
&&\frac{d^2r}{d\lambda^2} + 2\lf \frac{R_{,tr}-R_{,t}E_{,r}/E}{R_{,r}-RE_{,r}/E}\rg(\frac{dt}{d\lambda})(\frac{dr}{d\lambda}) \\ \nonumber
&&+ \Big{(} \frac{R_{,rr}-R_{,r}E_{,r}/E-RE_{,rr}/E+R(E_{,r}/E)^2}{R_{,r}-RE_{,r}/E}\\ \nonumber 
&&+\frac{k_{,r}}{2(1-k)}\Big{)} (\frac{dr}{d\lambda})^2 + 2(\frac{R}{E^2})\lf\frac{E_{,r}E_{,p}-EE_{,pr}}{R_{,r}-RE_{,r}/E}\rg(\frac{dr}{d\lambda})(\frac{dp}{d\lambda}) \\ \nonumber
&&+ 2(\frac{R}{E^2})\lf\frac{E_{,r}E_{,q}-EE_{,qr}}{R_{,r}-RE_{,r}/E}\rg(\frac{dr}{d\lambda})(\frac{dq}{d\lambda}) \\ \nonumber
&&- (\frac{R}{E^2})\lf\frac{1-k}{R_{,r}-RE_{,r}/E}\rg((\frac{dp}{d\lambda})^2+\lf\frac{dq}{d\lambda})^2\rg = 0 
\eea
\bea
&&\frac{d^2p}{d\lambda^2} + 2(\frac{R_{,t}}{R})(\frac{dt}{d\lambda})(\frac{dp}{d\lambda}) \\ \nonumber
&&-\Big{(}\frac{1}{R}\frac{R_{,r}-RE_{,r}/E}{1-k}(E_{,r}E_{,p}-EE_{,pr})\Big{)}(\frac{dr}{d\lambda})^2 \\ \nonumber
&&+ 2\Big{(}\frac{R_{,r}}{R}-\frac{E_{,r}}{E}\Big{)}(\frac{dr}{d\lambda})(\frac{dp}{d\lambda})- (\frac{E_{,p}}{E})(\frac{dp}{d\lambda})^2 \\ \nonumber
&&- 2(\frac{E_{,q}}{E})(\frac{dp}{d\lambda})(\frac{dq}{d\lambda})+ (\frac{E_{,p}}{E})(\frac{dq}{d\lambda})^2 = 0  \\ \nonumber
\label{eq:NGp}  
\eea
\bea
\label{eq:NGq} 
&&\frac{d^2q}{d\lambda^2} + 2(\frac{R_{,t}}{R})(\frac{dt}{d\lambda})(\frac{dq}{d\lambda})\\ \nonumber 
&&-\Big{(}\frac{1}{R}\frac{R_{,r}-RE_{,r}/E}{1-k}(E_{,r}E_{,q}-EE_{,qr})\Big{)}(\frac{dr}{d\lambda})^2 \\ \nonumber
&&+ 2\Big{(}\frac{R_{,r}}{R}-\frac{E_{,r}}{E}\Big{)}(\frac{dr}{d\lambda})(\frac{dq}{d\lambda})-(\frac{E_{,q}}{E})(\frac{dp}{d\lambda})^2 \\ \nonumber 
&&- 2(\frac{E_{,p}}{E})(\frac{dp}{d\lambda})(\frac{dq}{d\lambda})+ (\frac{E_{,q}}{E})(\frac{dq}{d\lambda})^2 = 0 \\ \nonumber
\eea
where $\lambda$ is a parameter. Next, the first integral (i.e. $k_ak^a=0$) reads   
\be
\small{
-(\frac{dt}{d\lambda})^2 + \frac{(R_{,r}-RE_{,r}/E)^2}{1-k}(\frac{dr}{d\lambda})^2 + \frac{R^2}{E^2}((\frac{dp}{d\lambda})^2+(\frac{dq}{d\lambda})^2) = 0 }.
\label{eq:FirstInt}
\ee 

In order to study the propagation of null rays in the Szekeres models, one need to integrate numerically the set of equations (\ref{eq:NGt})-(\ref{eq:NGq}) subject to equation (\ref{eq:FirstInt}). 

Before to do that, let's relate these equations to the redshift. For that, we consider two consecutive wavecrests of a light ray traveling to the observer.  One is described by the first integral at $t(\lambda)$ and the second is at $t(\lambda)+T(\lambda)$, where T is, therefore, the period of the light. Thus we have:
\be
\small{
(\frac{dt}{d\lambda})^2 = \frac{(R_{,r}-RE_{,r}/E)^2}{1-k}(\frac{dr}{d\lambda})^2 + \frac{R^2}{E^2}((\frac{dp}{d\lambda})^2+(\frac{dq}{d\lambda})^2)   }
\label{eq:FIt}
\ee
\bea
(\frac{d(t+T)}{d\lambda})^2 &= \frac{(R(t+T,r)_{,r}-R(t+T,r)E_{,r}/E)^2}{1-k}(\frac{dr}{d\lambda})^2 \\ \nonumber
&+ \frac{R^2(t+T,r)}{E^2}((\frac{dp}{d\lambda})^2+(\frac{dq}{d\lambda})^2)
\label{eq:FItT}
\eea
We can now use the first integral to see how the period changes as the light propagates.  We will assume that the period of the light, $T$, is much  smaller than the time over which the light propagates, see for example \cite{PlebanskiAndKrasinski2006}.  This allows us to Taylor expand to first order 
\be
R(t+T,r)= R(t,r)+\dot{R}(t,r)T
\ee
\noindent and
\be
R'(t+T,r)=R'(t,r)+\dot{R}'(t,r)T
\ee
\noindent substituting into (\ref{eq:FIt}) and (\ref{eq:FItT}), and subtracting (\ref{eq:FIt}) from (\ref{eq:FItT}) we get:
\bea
\frac{dt}{d\lambda}\frac{dT}{d\lambda} &= T(\lambda) ((\frac{\dot{R}'R+R\dot{R}(\frac{E'}{E})^2 
-(R'\dot{R}+R\dot{R}')\frac{E'}{E}}{1-k})(\frac{dr}{d\lambda})^2 \\ \nonumber
&+\frac{R\dot{R}}{E^2}((\frac{dp}{d\lambda})^2+(\frac{dq}{d\lambda})^2))
\label{eq:FIdtdT} 
\eea
Now, redshift is related to $T$ by :
\be
1+z(\lambda_e)=\frac{T(\lambda_o)}{T(\lambda_e)}
\ee
\noindent where the subscript $e$ is at emission and $o$ is at observation.  We take the derivative with respect to $\lambda_e$ to obtain:
\be
\frac{dz}{d\lambda_e}=-\frac{dT(\lambda_e)}{d\lambda_e}\frac{1+z(\lambda_e)}{T(\lambda_e)}.
\label{eq:diffz}
\ee
We can now substitute (\ref{eq:diffz}) into (\ref{eq:FIdtdT}) and drop the subscripts to get:
\bea
\nonumber \frac{dz}{d\lambda}&=-(1+z)(\frac{d\lambda}{dt}) ((\frac{\dot{R}'R+R\dot{R}(\frac{E'}{E})^2-(R'\dot{R}+R\dot{R}')\frac{E'}{E}}{1-k})(\frac{dr}{d\lambda})^2 \\ 
&+\frac{R\dot{R}}{E^2}((\frac{dp}{d\lambda})^2+(\frac{dq}{d\lambda})^2))
\eea
\noindent or more simply:
\bea
\nonumber \frac{d(ln(1+z))}{d\lambda}&=-\frac{1}{\frac{dt}{d\lambda}} ((\frac{\dot{R}'R+R\dot{R}(\frac{E'}{E})^2-(R'\dot{R}+R\dot{R}')\frac{E'}{E}}{1-k})(\frac{dr}{d\lambda})^2 \\ 
&+\frac{R\dot{R}}{E^2}((\frac{dp}{d\lambda})^2+(\frac{dq}{d\lambda})^2))
\label{eq:redRel}
\eea


Now, equations (\ref{eq:NGt})-(\ref{eq:NGq}) constitute a system of second-order ordinary differential equations (ODEs) for the functions $\{t(\lambda), r(\lambda), p(\lambda), q(\lambda)\}$ where the coefficients are composed of the metric functions evaluated on the null cone using the Field equations and the model specifications. Equation ({\ref{eq:redRel}) relates the system to the redshift. 

We integrate the system using a fourth-order Runge-Kutta algorithm with adaptive step size \cite{Press}. Our numerical code iterates between calls to evaluate the Field equations on the null cone and calls to integrate the ODEs. We also coded a 3-dimensional interpolator \cite{Press} in order to select geodesics with starting points matching the coordinates of the supernovae of interest as given in \cite{SupernovaDatabase}. We implemented the Runge-Kutta code \cite{Press} with the function vectors 
\be
\textbf{y}=\{t,r,p,q,\frac{dt}{dl},\frac{dr}{dl},\frac{dp}{dl},\frac{dq}{dl}\}
\ee
and 
\be
\frac{d\textbf{y}}{dl}=\{\frac{dt}{dl},\frac{dr}{dl},\frac{dp}{dl},\frac{dq}{dl}\,\frac{d^2t}{dl^2},\frac{d^2r}{dl^2},\frac{d^2p}{dl^2},\frac{d^2q}{dl^2}\}
\ee
so that the system of 4 second-order ODEs is reduced into a system of 8 first-order ODEs. We also use the parameter $l\equiv \ln (1+z)$ instead of $\lambda$. Further detail about the numerical integration and the 3D-interpolator will be presented in a separate follow up paper \cite{Ishaketal2008}.  

The next step of our numerical integration is to evaluate numerically the area distance and the luminosity distance. For a general definition of the observer area distance, $r_0$, we go back to earlier work by \cite{Ellisetal1985,Stoegeretal1992} where for a general metric in spherical coordinates, $r_0$ is given by 
\be
r_0^4 \sin^2\theta =  \, {\rm det} (H_{_{IJ}})
\label{eq:area_distance}
\ee 
where $H_{_{IJ}}$ is the $\theta$-$\phi$ block of the metric. For that, it is useful to rewrite the Szekeres metric in spherical coordinates. For this purpose we consider the following coordinate transformation:
\bea
p  &=&  \,{\rm cot} \left( \frac{ \theta }{2} \right) \cos (\phi), \nonumber \\
q  &=&  \,{\rm cot} \left( \frac{ \theta }{2} \right) \sin (\phi), \nonumber \\
r &=&  r,
\label{eq:coordTrans}
\eea
that yields the new line element
\be
ds^2= -dt^2+\frac{(R'-R\frac{\tilde{E'}}{\tilde{E}})^2}{\epsilon-k(r)}dr^2 +\big{(}\frac{R}{\tilde{E}}\big{)}^2({d\theta^2+\sin^2(\theta) d\phi^2})
\label{eq:SzekeresSS}
\ee
where the function $\tilde{E}\equiv E (1-\cos(\theta))$ does not alter other parts of the metric. 

Now, we note from the metric (\ref{eq:SzekeresSS}) that there are no off--diagonal components of the $\{\theta,\phi\}$ part of the metric, i.e $g_{\theta\phi}=0$, and since we specialize here our analysis to the case of Szekeres models with $\epsilon=+1$, then, by (\ref{eq:area_distance}), the area distance can be described by \cite{AreaDistance}
\be
r_0 = \frac{R}{E}.
\ee
which indeed depends on the three coordinates $r$, $\theta$ and $\phi$ on the null cone. It follows that the luminosity distance is also a function of the three coordinates and is given by 
\be
d_L(z)=(1+z)^2\, \frac{R}{E}.
\label{eq:ldistance}
\ee
We integrate numerically $d_L$ along with the system of ODEs (\ref{eq:NGt})-(\ref{eq:NGq}) above. Then, as usual, we evaluate the distance modulus to supernova as  
\be
m(z)-M=5 \log (d_L) + 25
\ee
and we use the supernova magnitudes and uncertainties from the combined data sets of \cite{Davisetal2007,WoodVasey2007,Riessetal2007}  and we transform the coordinates from the supernova database \cite{SupernovaDatabase}. The findings are presented in the next section. 
{
\begin{figure}[t]
\begin{center}
\includegraphics[width=2.4in,height=3.1in,angle=-90]{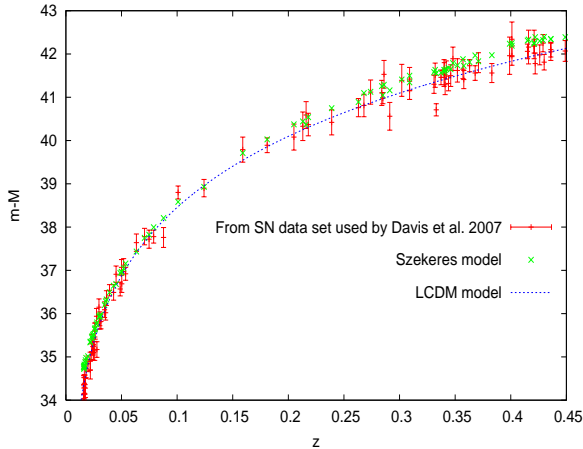}
\caption{\label{fig:hubble} 
Supernova fits for the Szekeres model (green crosses) and LCDM (blue curve) models. The data is 94 Supernova (up to $1+z=1.449$) from Davis et al 2007, Wood-Vasey et al 2007, and Riess et al 2007 \cite{Davisetal2007,WoodVasey2007,Riessetal2007}. 
The Szekeres model fits the data with a $\chi^2=112$. This is close to the $\chi^2=105$ of the LCDM concordance model. Because of the possible systematic uncertainties in the supernova data, it is not clear that the difference between the two $\chi^2$ is significant. Furthermore, other Szekeres models remain to be explored. The Szekeres model used is also consistent with the requirement of spatial flatness at CMB scales.}
\end{center}
\end{figure}
}

\section{results and discussion}
We discuss here our results from the numerical integration and comparison to combined supernova data sets of \cite{Davisetal2007,WoodVasey2007,Riessetal2007}.
We specify our Szekeres model by using some guidance from current observations of matter abundances, CMB observations, and also some previous works on inhomogeneous models \cite{Bolejko2006a,Bolejko2007,Garfinkle2006}.  
\begin{itemize}
  \item {We consider the case of Szekeres models with $\epsilon=+1$. This case includes the three sub-cases of hyperbolic, elliptic, and parabolic models.}
\item {We set $t_B(r)=0$ for a simultaneous Big Bang like in the FLRW models.}
  \item {In the usual way, e.g. \cite{HellabyAndKrasinski2002}, we use the coordinate freedom in $r$ and set $M(r)=(\sinh(r))^3$.}
  \item {In a similar way to the work done on the LTB models by \cite{Garfinkle2006}, we set here, for the Szekeres models $k(r)=\tilde{k}(r)r^2$ with 
\be
\tilde{k}(r)=\frac{-1}{1+ C^2 r^2}. 
\label{eq:curvature}
\ee
So at large $r$, the spatial curvature goes to zero as indicated by CMB observations \cite{observations5,observations9} and at very small $r$ curvature goes negative in accord with local observations of matter abundances with a density about one quarter of the critical density. The constant $C$ will be set by best fit model to the data.}
  \item {For $\{S,P,Q\}$, we use model-1 of \cite{Bolejko2006a} with $\{140,10, -113\,ln(1+r)\}$ that was considered within the context of structure formation using Szekeres and also within a general discussion of cosmological applications of the models \cite{Bolejko2006b}. Other sub-models will be explored in forthcoming investigations.}
  \end{itemize}

Our fitings are presented in FIG.~1. We find that the Szekeres model that we considered fits the supernova data competitively up to a redshift $1+z=1.449$. The best fit model has a value $C=0.80$ with a $\chi^2=112$, and this is found to be close to the $\chi^2=105$ of the flat LCDM model with $\Omega_m=0.27$ and $\Omega_\Lambda=0.73$ (see also \cite{Davisetal2007} for the LCDM model). In view of the possible systematic uncertainties involved in the supernova data, it is not clear that the difference between the two $\chi^2$ is significant. Also, the model that we used here was derived and discussed from fits that modeled large scale structure \cite{Bolejko2006a} (clusters and voids), so while the model is found to provide a good fit to the supernova data with ($1+z<1.45$), one needs to take into account the transition to larger scales. Indeed, it is well known that the Szekeres models average to almost Friedmann models at very large scale of distances \cite{Szekeres1,Szekeres2,Szafron1977,BonnorAndTomimura1976,Bonnoretal1977,GoodeAndWainwright1982,Bonnor1985,HellabyAndKrasinski2002} but further work is needed in our numerical approach in order to model such a transition and include it into the fit. Nevertheless, as shown in equation (\ref{eq:curvature}), the constant $C$ controls spatial curvature and our model is consistent with the requirement of spatial flatness at the CMB scales.

Now, while the LCDM model requires a cosmological constant in order to fit the data, the Szekeres model is used without any dark energy component. As indicated by the function $\tilde{k}(r)$ in (\ref{eq:curvature}), the interpretation of this result is that we may happen to live in one of the many under-dense regions of the universe as described by a Szekeres inhomogeneous cosmological model and this would leads to an apparent acceleration.  

However, a number of points need to be investigated in order to fully explore this possibility and work is ongoing by the authors. These include:
\\
$\bullet$ {Exploration of other sub-models and ansatz for the functions S, P, Q and the curvature function $k(r)$ of the model including the transition from Szekeres models to an almost FLRW models}
\\
$\bullet$ {Comparison to CMB observables. We expect that the position of the first peak of the CMB spectrum can be easily met by the Szekeres model that we considered. Most importantly, the models show promise to fit the full CMB spectra because of their flexible geometrical structure. Another constraint that need to be studied is the amplitude of primordial fluctuations in the gravitational field. Here, one needs to use the Szekeres-Szafron models \cite{Szafron1977,SussmanAndTriginer1999} that include pressure as well. }
\\
$\bullet$ {Comparison of the models to observables of large-scale structure growth, such as gravitational lensing and galaxy clustering}
\\
\indent 

In conclusion, this first work of a series shows that the Szekeres inhomogeneous models provide a good fit to the supernova data and seems to support apparent acceleration but further investigations using these models are required in order to thoroughly explore this possibility. 
\acknowledgments
We thank Krzysztof Bolejko for useful comments on the paper. We thank Eric Linder, James Peebles, Joseph Silk, Adam Riess, and Ned Wright for useful correspondence on this topic. We thank Tamara Davis and Adam Riess for useful comments about the supernova data sets.
M.I. acknowledges partial support from the Hoblitzelle Foundation and a Clark award. 
{}

\begin{thebibliography}{}
\bibitem{AreaDistance}
The area of the ($t=const$, $r=const$) surfaces is finite in the $\epsilon=+1$ case that we consider here while it can be infinite in the cases $\epsilon=0$ and $\epsilon=-1$ \cite{HellabyAndKrasinski2002,HellabyAndKrasinski2008}.
\end{thebibliography}
\end{document}